\documentclass[11pt, letterpaper]{article}
\usepackage[utf8]{inputenc}
\usepackage[margin=1in]{geometry}
\usepackage{titlesec}
\usepackage{tabu}
\usepackage{enumitem}
\usepackage{amssymb}
\usepackage{graphicx,amsmath}
\usepackage[square,sort,comma,numbers]{natbib}
\usepackage{hyperref}

\newlist{selectlist}{itemize}{2}
\setlist[selectlist]{label=$\square$,leftmargin=*,noitemsep,topsep=0pt}

\titleformat{\section}[block]{\hspace{1em}\bfseries}{\thesection.}{0.5em}{} 
\titleformat{\subsection}[block]{\hspace{1em}}{\thesubsection}{0.5em}{}

\begin{document}
\begin{flushleft}

\setlength{\parindent}{0pt}
\setlength{\parskip}{10pt}

\textbf{Title:} Methods to simplify cooling of liquid Helium cryostats

\textbf{Authors:} Rafael \'Alvarez Montoya$^1$, Sara Delgado$^1$, Jos\'e Castilla$^1$, Jos\'e Navarrete$^2$, Nuria D\'iaz Contreras$^2$, Juan Ram\'on Marijuan$^2$, V\'ictor Barrena,$^1$, Isabel Guillam\'on$^1$ and Hermann Suderow$^1$

\textbf{Affiliations:} $^1$ Laboratorio de Bajas Temperaturas, Departamento de
F\'isica de la Materia Condensada \\ Instituto de Ciencia de
Materiales Nicol\'as Cabrera and Condensed Matter Physics Center (IFIMAC), Facultad de Ciencias \\ Universidad Aut\'onoma de Madrid, 28049 Madrid, Spain\\ $^2$ Segainvex \\ Universidad Aut\'onoma de Madrid, 28049 Madrid, Spain

\textbf{Contact email:} hermann.suderow@uam.es

\textbf{Abstract:} Liquid Helium is used widely, from hospitals to characterization of materials at low temperatures. Many experiments at low temperatures require liquid Helium, particularly when vibration isolation precludes the use of cryocoolers and when one needs to cool heavy equipment such as superconducting coils. Here we describe methods to simplify the operations required to use liquid Helium by eliminating the use of high pressure bottles, avoiding blockage and improving heating and cooling rates. First we show a simple and very low cost method to transfer liquid Helium from a transport container into a cryostat that uses a manual pump having pumping and pressurizing ports, giving a liquid Helium transfer rate of about 100 liters an hour. Second, we describe a closed cycle circuit of Helium gas cooled in an external liquid nitrogen bath that allows precooling a cryogenic experiment without inserting liquid nitrogen into the cryostat, eliminating problems associated to the presence of nitrogen around superconducting magnets. And third, we show a sliding seal assembly and an inner vacuum chamber design that allows inserting large experiments into liquid Helium. 

 Materials characterization at low temperatures and high magnetic fields.

\newpage

\textbf{Specifications tables:}

\tabulinesep=1ex
\begin{tabu} to \linewidth {|X|X[3,l]|}
\hline  \textbf{Hardware name} & Helium transfer system based on manual pump
  \\
  \hline \textbf{Subject area} & %
  \begin{itemize}
  \item Materials Science
  \item Physical property measurements
  \item Cryogenics of magnetic resonance equipment
  \end{itemize}
  \\
  \hline \textbf{Hardware type} &
  \begin{itemize}
  \item Device to transfer liquid Helium
  \item Maintenance and preparation of systems for physical property measurements at low temperatures
  \end{itemize}
  \\ 
\hline \textbf{Open source license} &
  CERN OHL
  \\
\hline \textbf{Cost of hardware} &
  30 \$
  \\
\hline \textbf{Source file repository} & 
  https://osf.io/e6k7r/
\\\hline
\end{tabu}
 
\tabulinesep=1ex
\begin{tabu} to \linewidth {|X|X[3,l]|}
\hline  \textbf{Hardware name} & Closed cycle Helium gas precooling system
  \\
  \hline \textbf{Subject area} & %
  \begin{itemize}
  \item Materials Science
  \item Physical property measurements
  \end{itemize}
  \\
  \hline \textbf{Hardware type} &
  \begin{itemize}
  \item Precooling equipment to liquid nitrogen temperatures
  \item Physical property measurements at low temperatures
  \end{itemize}
  \\ 
\hline \textbf{Open source license} &
  CERN OHL
  \\
\hline \textbf{Cost of hardware} &
  1500 \$
  \\
\hline \textbf{Source file repository} & 
  https://osf.io/e6k7r/
\\\hline
\end{tabu}

\newpage

\tabulinesep=1ex
\begin{tabu} to \linewidth {|X|X[3,l]|}
\hline  \textbf{Hardware name} & Sliding seal and inner vacuum chamber for rapid sample exchange
  \\
  \hline \textbf{Subject area} & %
  \begin{itemize}
  \item Materials Science
  \item Physical property measurements
  \end{itemize}
  \\
  \hline \textbf{Hardware type} &
  \begin{itemize}
  \item Precooling equipment to liquid nitrogen temperatures
  \item Physical property measurements at low temperatures
  \end{itemize}
  \\ 
\hline \textbf{Open source license} &
  \textit{CERN OHL}
  \\
\hline \textbf{Cost of hardware} &
  \textit{200 \$}
  \\
\hline \textbf{Source file repository} & 
  https://osf.io/e6k7r/
\\\hline
\end{tabu}

\end{flushleft}

\section{Hardware in context}

Much of the innovative work in cryogenics is now devoted into dry or closed-cycle cryogenic systems, driven by the development of thermoacoustic systems and the improvement of thermodynamic machines \cite{Radebaugh1990}. However, the improvements have not displaced liquid Helium, but considerably facilitated its development and use in different fields. For example, magnetic resonance imaging cryostats include coolers providing improved shielding that nearly zero the consumption of liquid Helium\cite{Ackermann2002,recondens}. In research laboratories, large systems including superconducting magnets require most often liquid Helium, because cooling using the huge enthalphy of Helium gas is unmatched with respect to any other cooling method. In addition, there is an increasing need for experiments in which the mechanical vibrations are held to a minimum, for which the use of liquid Helium is unavoidable\cite{doi:10.1063/1.1149605,doi:10.1063/1.3520482,doi:10.1063/1.2804165,doi:10.1063/1.4999555,doi:10.1063/1.4822271,doi:10.1063/1.4793793,doi:10.1063/1.4769258,doi:10.1063/1.5049619}. Therefore, research in wet cryogenics is needed to improve the performance of systems using liquid Helium.

In particular, cooling of large cryostats bears some inherent problems related to the manipulation of liquid Helium. There are many laboratory books and notes that are there to help the users, in addition to usual manuals provided by companies delivering cryostats \cite{hitchhiker,Oxford}. Most of these manuals explain transfer of liquid Helium and sample turn-around methods. But several problems are repeatedly mentioned. These are about handling of liquid Helium bottles and transfer methods, inserting and cooling to liquid Helium temperatures large inserts, often containing a dilution refrigerator unit, and precooling to liquid nitrogen temperatures. For example, as we explain below, tubes can be blocked when liquid nitrogen has been used and not fully removed before transferring liquid Helium, or Helium leaks appear when inserting large systems into liquid Helium.

Here we discuss possible solutions to these issues, which we developed in the use of a set of five large dilution refrigerator units that are equipped with Scanning Tunneling Microscopes. The microscopes have been partially described in detail in Ref.\cite{doi:10.1063/1.3567008}. A cryogenic set-up with a three axis vector magnet and associated electronics for the magnet and the microscope has been described in Ref.\cite{doi:10.1063/1.4905531}. The solutions described here have been developed more recently. The Helium transfer manual pump has not been reported till now, to our knowledge. The liquid nitrogen precooling device has been reported in Ref.\cite{cooling} in a realization that includes a specific compressor and heat exchanging system. Here we show a much simpler realization, using commercially available equipment. The improvements in sliding seal assembly and vacuum chamber have not been reported to our knowledge.

\section{Hardware description}


Potential advantages of using the described hardware:

\begin{itemize}
\item Eliminating high pressure Helium equipment from cryogenic laboratories.
\item Avoiding blockage and problems with transfer of liquid nitrogen into cryostats.
\item Improving sample turn around time in cryogenic devices.
\end{itemize}

\subsection{Helium transfer system based on manual pump}

Liquid Helium is usually transferred from a transport Dewar to a cryostat using a transfer tube and applying an overpressure to the transport Dewar. An acceptable flow rate is usually of about 50 liters of liquid Helium for half an hour, which makes a volumetric flow rate $Q\approx 30$ ml/s. With a cross section of a pipe which typically is of about $d=10 mm^2$, this gives a velocity of liquid Helium close to the speed of sound. The force needed to produce Helium flow is the force needed to overcome height differences in the liquid Helium levels of both containers and to overcome friction. An overpressure of 0.2-0.4 bar in a container of typically a quarter of a square meter cross section is enough to transfer liquid Helium into the cryostat. Although the Reynolds number can be as high as $10^6$, losses due to friction do not hamper significantly the transfer process\cite{SKRBEK2004354}. The high Reynolds number makes liquid Helium flow a rather unusual process, which is however routinely carried out in laboratories.

To produce the overpressure of, say, 0.5 bar, one can simply introduce Helium gas into the transport Dewar by some means. A few simple calculations show that the evaporation produced in the liquid by the arrival of a small amount of hot gas is by far enough to produce an overpressure in the transport Dewar. For example, in a usual 100 liters transport Dewar, there is somewhat less than a $m^3$ room for the gas. The latent heat of vaporization of liquid Helium is of about 20kJ/kg. The enthalpy of Helium gas between room temperature and liquid Helium temperatures is of about 1.5 MJ/kg. Cooling of a kg of gas to liquid Helium temperatures requires evaporating about 75 kg of liquid. Taking into account the different densities, this implies that to achieve the required overpressure, one just needs to evaporate about a liter of liquid and that this can be done by cooling to liquid Helium of just about 12 liters of gas. Therefore, one needs to insert into the transport Dewar a small amount of Helium gas at room temperature to produce the required overpressure.

\begin{figure}[ht]
\includegraphics[angle=0,width=\textwidth,clip]{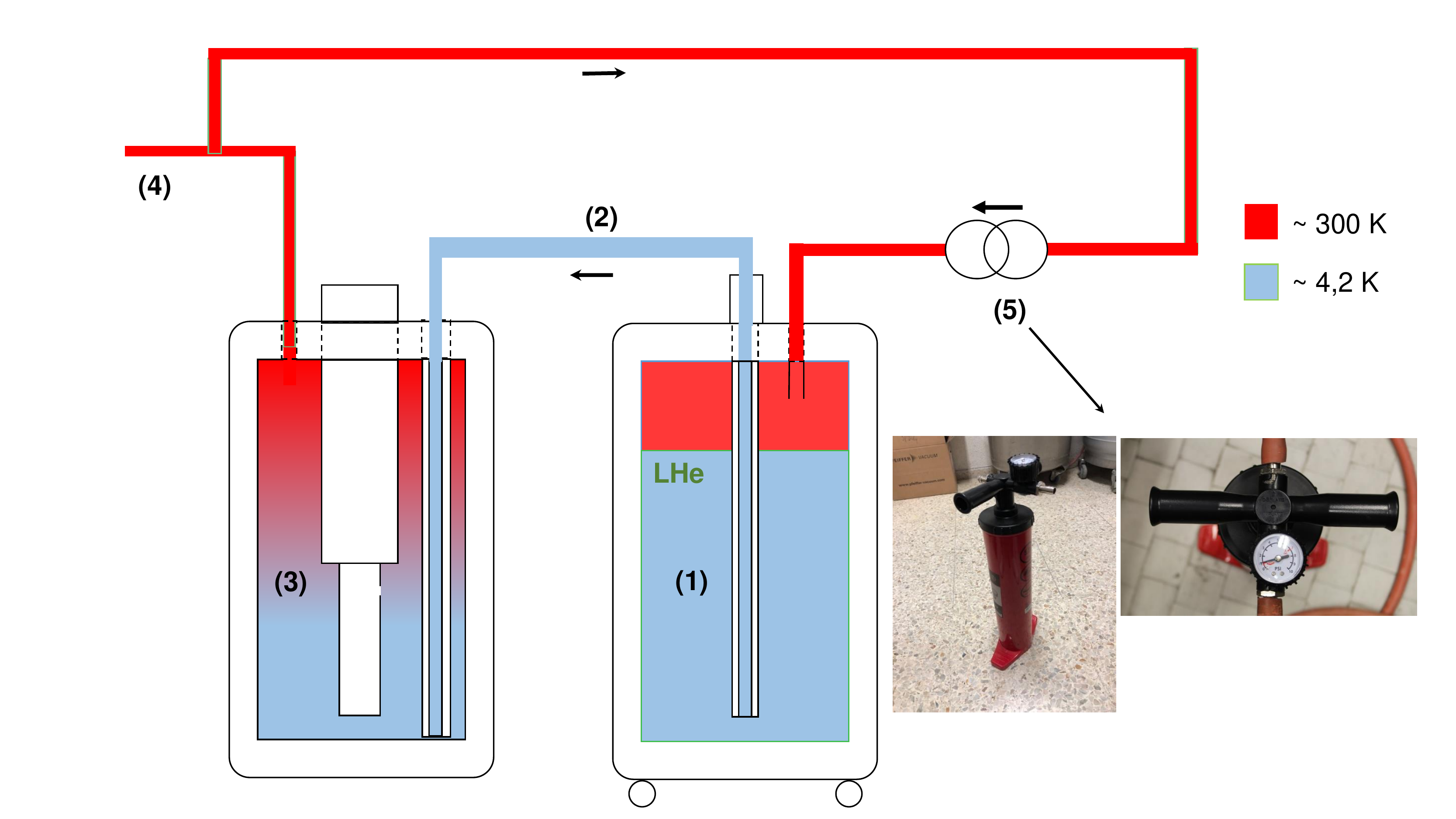}
\vskip -0cm
\caption{Scheme of the liquid Helium transfer system. We show schematically a liquid Helium transport Dewar (1), a transfer tube (2) and a cryostat (3). The cryostat is connected to the Helium recovery line (4). The transport Dewar (1) is connected to the recovery line (4) through a manual pump (5). The manual pump (5) is shown as a photo in the lower right part of the figure. The manual gas pump has clear entry and exit ports. Connectors for rubber tubes are glued to the ports. About five strokes of the manual pump are needed to fastly send Helium gas into the Dewar. The Helium gas is cooled by the liquid Helium, which evaporates and increases the pressure in the transport Dewar. The evaporation produces a small overpressure of about 0.5 bar, with which liquid Helium is transferred to the cryostat. Using this pump, 100 liters can be transferred in about an hour.} \label{Fig1}
\end{figure}

A usual solution is a small laboratory membrane compressor, which allows in principle for automated operation, but it is often not very practical. The usual flow rate of a small devices is of about 10 liters per minute or less. In practice, the compressor is often stuck by impurities or debris from the recovery line and one needs to operate the compressor for several minutes to reach the required pressure. This implies that the gas has enough time to thermalize inside the transport Dewar and remain at the upper level, without really evaporating the liquid. Furthermore, care has to be taken to make sure that Helium is not pumped out of the recovery line when it is not needed, which would imply losses of Helium gas. Therefore, most laboratories use insted Helium gas bottles pressurized to 200 bar with a pressure reducing manometer. However, high pressure gas bottles requires significant security management and it is quite easy to loose large amounts of gas by bad manipulation.

Rubber bladders can be used when slow filling rates are required\cite{Oxford}. The bladder is inflated with the small overpressure of the Helium Dewar. By pushing on the bladder, a few liters of Helium gas are moved inside the Dewar and the obtained flow of gas evaporates a small amount of liquid, which leads to an increase in pressure. However, the bladder is inflated and it becomes very difficult to manipulate. Small pressures can be nicely regulated this way, but when high transfer rates are needed, the bladder becomes too big to operate.

Recently, manual gas pumps with an entry and an exit port have been made available in the market. These are used to fill and empty large containers with air, such as boats or camping appliances. They pump gas volumes between 5 and 10 liters per stroke and a maximal pressure difference of order of 0.5 bar (see for example \url{https://www.decathlon.co.uk/52-l-hand-pump-id_8243066.html}). We have adapted one of such pumps to collect Helium from the recovery of the cryostat and insert the gas into the transfer Dewar. The pump has the speed and size to move enough Helium gas inside the transport Dewar and produce the required liquid Helium evaporation. We show schematically in Fig.\ref{Fig1} the whole arrangement. A few strokes to the pump are enough to produce the required increase in the pressure of the transport Dewar and fill a cryostat to with a velocity of about 100 liters an hour.

\subsection{Closed cycle Helium gas precooling system}

As mentioned above, the latent heat of vaporatization of liquid Helium is small (21 kJ/kg), so that precooling the system to liquid nitrogen allows reducing considerably the Helium boil-off. On the other hand, the melting point of liquid nitrogen is just at 63 K, quite close to 77 K, and the latent heat of fusion is of about 26 kJ/kg. This implies that, if the liquid nitrogen is not fully removed when the system has cooled, a sizeable amount of liquid Helium is needed to solidify the liquid nitrogen and cool down to liquid Helium temperatures. For example, if just an amount of liquid nitrogen of order of a cup, or of about 100 g of liquid nitrogen, is left in the system, 2.6 kJ are needed to solidify it, which implies evaporating about a kg, or about 10 l, of liquid Helium.

Removing all liquid nitrogen is not easy. One can use a tube inserted down to the bottom of the cryostat and an overpressure, so that nitrogen is transferred back outside the cryostat. However, the tube must be designed in such a way as to reach the bottom of the cryostat so that all nitrogen can be removed. In addition, the cryostat must be flushed and pumped several times\cite{Oxford}. There is always a risk of leaving small amounts of nitrogen in thin tubes, and this can lead to blockage when cooling to liquid Helium. Furthermore, nitrogen might get trapped inside superconducting magnets, which produces strain that might lead to quench during operation.

\begin{figure}[ht]
\includegraphics[angle=0,width=0.9\textwidth,clip]{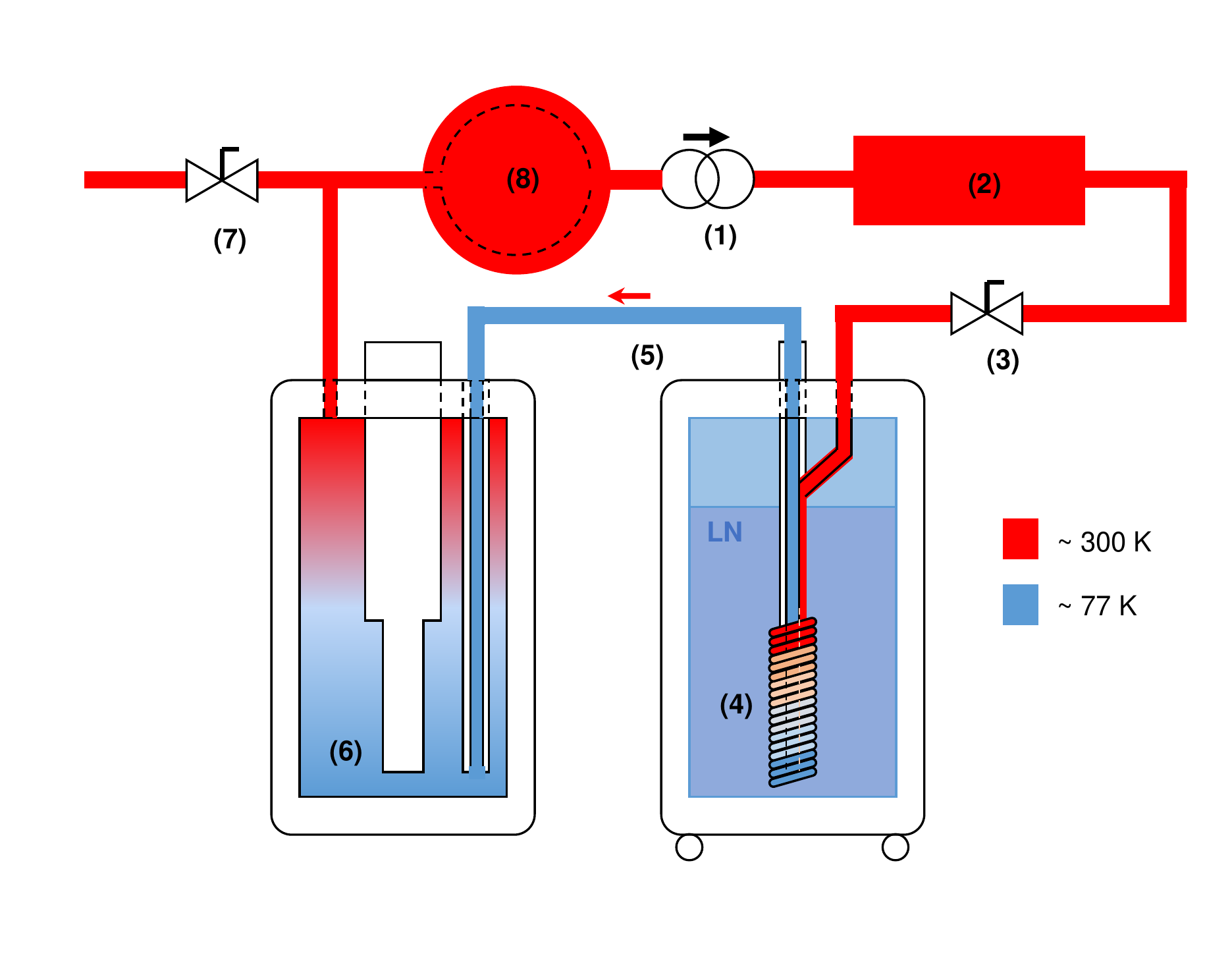}
\caption{Scheme of the precooling system. A compressor (1) is connected to a pressure container (2). A tube connects the container (2), through a regulating valve (3) to a heat exchanger immersed into liquid nitrogen (4). Helium flows out of the heat exchanger through a usual Helium transfer tube (5) into the cryostat (6). A valve (7) connects the recovery of the cryostat to the Helium recovery line of the laboratory. This valve is closed during operation and can be used to fill the whole circuit with Helium gas. The recovery of the cryostat is connected to a balloon (8) that reduces the time needed to operate the compressor (1) and allows for continous flow.} \label{Fig2}
\end{figure}

To avoid these difficulties, we have designed a precooling system as schematically described in Fig.\ref{Fig2}. Helium gas is circulated within a close circuit through a heat exchanger in liquid nitrogen, then inserted into the cryostat and flows again into the circuit after cooling the cryostat. We force Helium gas at room temperature through a serpentine immersed in liquid nitrogen and insert the outcoming cold gas using a usual Helium transfer tube into a cryostat with a superconducting coil inside. The warm Helium gas exiting the cryostat is again inserted into the circuit.

\begin{figure}[ht]
\includegraphics[angle=0,width=\textwidth,clip]{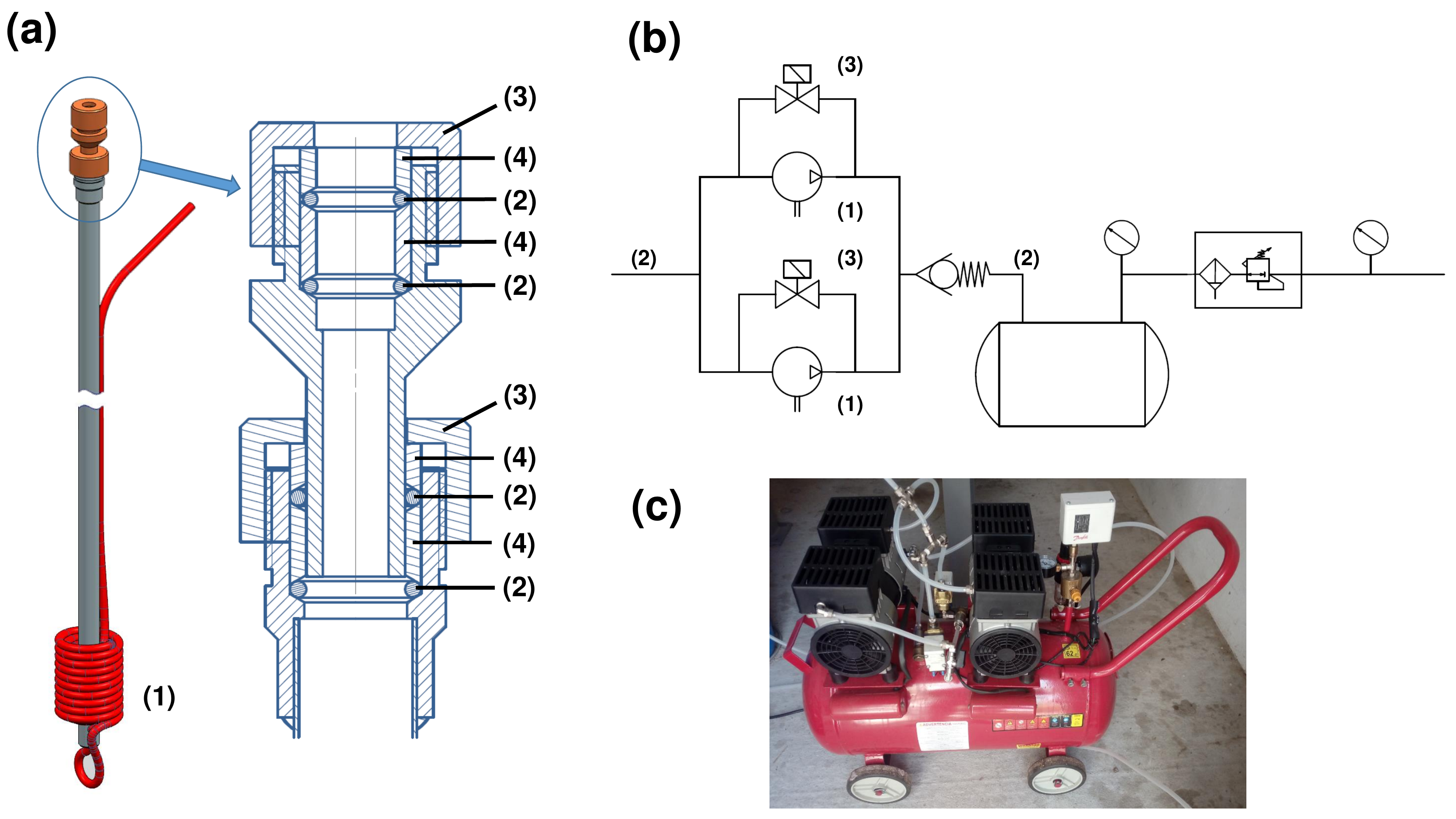}
\caption{(a) Scheme of the heat exchanger. The heat exchanger consists of a long copper tube wound at the bottom (1) in such a way that the external diameter enters a small liquid nitrogen container. The top of the heat exchanger (the exit) has a O-ring system that allows inserting transfer tubes of the two most common diameters (9 mm and 12 mm). It consists of four O-rings (2) that are held by threaded brass caps (3) and internal brass caps (4) that are machined at an angle at the ends to compress the O-ring to the transfer tube that is to be inserted. (b) Scheme of the compressor. The system (1) is a commercially available set of two membrane compressors mounted in parallel. The entry and exit ports (2) have been sealed using Araldite. The valves (3) have been exchanged by leak-tight vales and we have added a system that allows equilibrating pressures before re-starting the compressor. (c) Photography of the modified compressor with pressure container.} \label{Fig3}
\end{figure}

The same method has been proposed in the past\cite{cooling}, with, however, a rather complex heat exchanger and a multiple stage blade compressor. 

To discuss the working principle of the system, let us consider an example with a system whose cooling would be equivalent to cool about 50 kg of copper. The heat of vaporization of liquid nitrogen is of about 200 kJ/kg. Density of liquid nigrogen is of about 800 kg/m$^3$, so the vaporization of a m$^3$ of liquid provides 160MJ. Cooling 50 kg of copper from room temperature down to liquid nitrogen requires about 3 MJ (63 kJ per kg of copper), or 20 liters of liquid nitrogen.

On the other hand, Helium gas has a specific heat at constant volume of about 3kJ/kgK. Taking a density of 0.17 kg/m$^3$, this gives 0.5 kJ/m$^3$K. Assuming that Helium follows the ideal gas law and the specific heat is constant with temperature, we estimate that cooling Helium gas from room temperature to liquid nitrogen requires removing 100 kJ per m$^3$ of gas. If we use liquid nitrogen to cool the gas, we need about 0.6 liters of liquid nitrogen to cool a m$^3$ of Helium gas to liquid nitrogen temperatures.

To cool the amount of copper mentioned above (50 kg) to liquid nitrogen, we need to evaporate at least 20 liters of liquid nitrogen. Here we do this by using Helium gas that has been previously thermalized in the liquid nitrogen. This implies that we need to circulate about 30 m$^3$ of Helium gas through liquid nitrogen. Of course, best is to use a closed-cycle circuit. With a circulation of 10m$^3$/h of Helium gas through the liquid nitrogen, the required cooling is achieved within some hours. This circulation rate can be readily achieved using a medium size commercial membrane compressor.

It is relatively easy to build a heat exchanger (Fig.\ref{Fig3}(a)) that allows cooling such an amount of Helium gas with liquid nitrogen. A one-dimensional heat exchanger consisting of a 1 m long tube copper tube with a diameter of the order of a cm suffices to achieve the desired cooling.

The circuit requires an expansion Helium gas by the pressure difference which is established between the input in the cryostat and the exit. The Joule-Kelvin coefficient of Helium gas is small, of about 0.06 K/bar\cite{PhysRev.43.60}, so that the expansion implies a minimal heating (about 60 mK).

We have used this system to cool a cryostat with a superconducting coil and a dilution refrigerator and could reach temperatures of about 100 K overnight (see validation results in Fig.\ref{Fig5}) and then immediately started transferring Helium.

An relevant aspect is that, if the compressor is operated all the time, it might burn down the motor of a simple commercial device. Therefore, the circuit is not a simple flow through a compressor on a closed-cycle (Fig.\ref{Fig3}(b)). The pressure difference established by the flow of Helium gas occurs at the transfer tube (Fig.\ref{Fig2}) mostly. Using Hagen-Poiseuille, we obtain about 1 bar considering a transfer tube of 2 mm diameter and 3 m length. To avoid using the compressor all times, we insert a pressurized container with a valve limiting the flow rate (Fig.\ref{Fig3} and (2) at Fig.\ref{Fig2}). The compressor is activated whenever the pressure reaches 1.5 bar and stopped when it reaches 4 bar at the pressurized container. A valve at the exit of the pressurized container limits the pressure on the transfer line so that it remains below about 2 bar ((3) at Fig.\ref{Fig2}). In order to have enough gas to establish the overpressure each time the compressor is switched on again, we require a Helium balloon on the low pressure side ((8) at Fig.\ref{Fig2}). The balloon should be at least four times the volume of the pressurized container.

We used a container of 50 liters and a balloon of about 500 liters. When the gas flows out of the pressurized container, it slowly fills the balloon and when the pressure of the container reaches 1.5 bar, the compressor takes the gas from the balloon and fills the pressurized container.

The compressor worked in periods of about 2 minutes each 5 minutes, so that it remained within the specifications of the motor. The compressor used is a simple commercial membrane compressor. The scheme of the compressor is shown in Fig.\ref{Fig3}(b). We replaced the factory tubes by PVC pipings with pressure fittings. The compressor is activated by a pressure sensor installed in the container, that starts the motor at 1.5 bars and stops it at 4 bar. In addition, we installed two electrovalves that allow evacuating the overpressure in the exit of the compressor without emptying the container. This is needed as membrane compressors do not start with an overpressure at the exit. At the exit of the container, a flow regulator allows to establish the needed flow of gas through the circuit. A small filtering unit serves to remove impurities from the gas. A safety valve at 6 bar provides security for overpressure.

\subsection{Sliding seal assembly and inner vacuum chamber for rapid sample exchange}

\begin{figure}[ht]
\includegraphics[angle=0,width=\textwidth,clip]{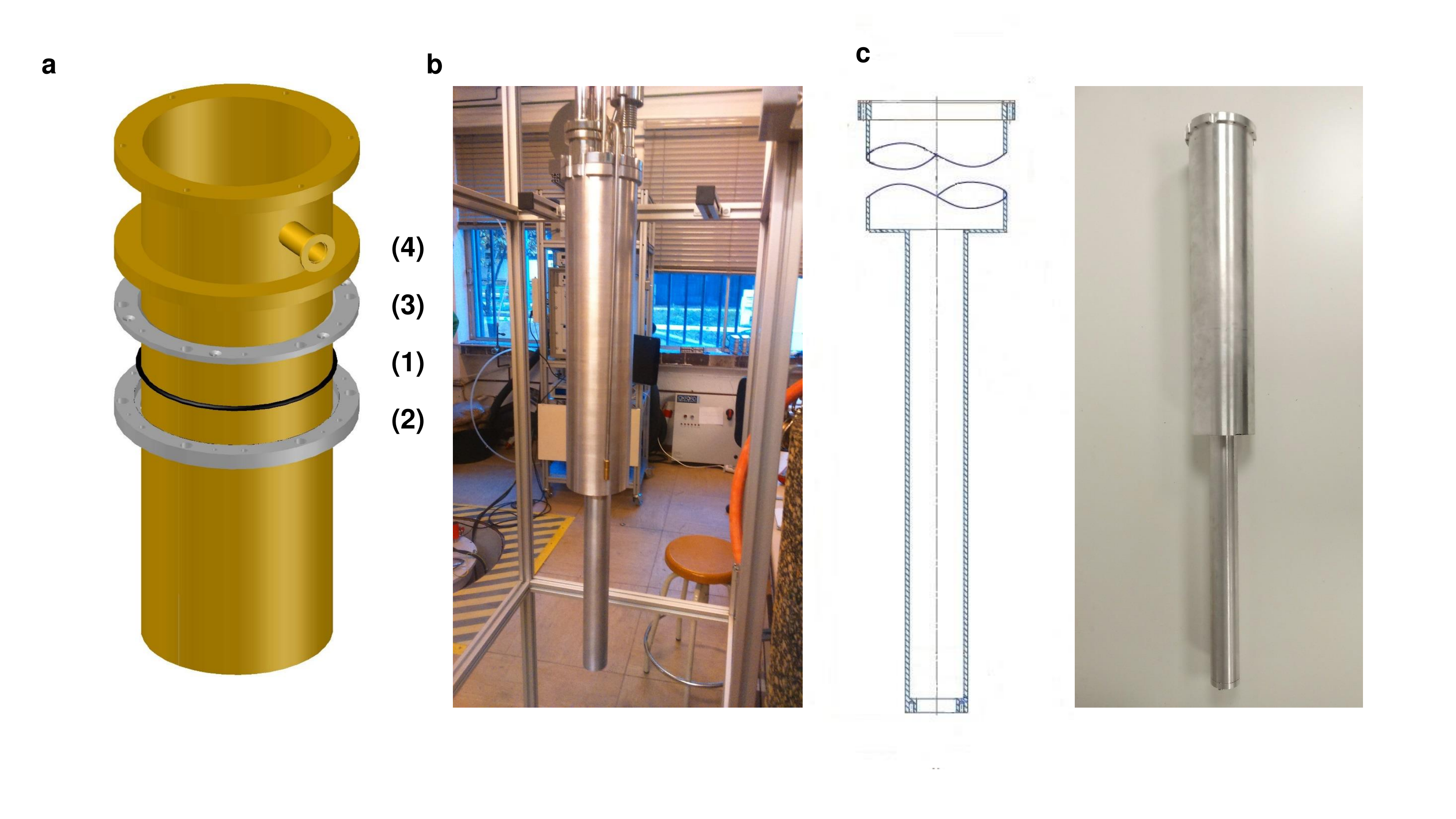}
\vskip -0cm
\caption{(a) Schematics of the sliding seal assembly consisting of an O-ring (1) and a bottom stainless steel piece (2) and a cap (3). The bottom part has threaded holes that allow fixing (3) in such a way as to compress the O-ring and additional holes allowing to fix the insert into the cryostat. (b) Photo of the insert outside and loaded into the cryostat. (c) Schematic drawing and photo of an Aluminium inner vaccum chamber.} \label{Fig4}
\end{figure}

Often, a refrigerator with diameter well above 100 mm has to be inserted into liquid Helium (see e.g. \cite{OxfordDilFridge}). This can be made by a system that allows precooling the assembly using Helium gas. To this end, the cryostat needs a cylindrical enclosure with a Helium recovery port and a sliding seal that blocks the vertical motion of the cryostat, particularly if an overpressure appears in the cryostat, and is at the same time leak tight to avoid loss of Helium gas. A solution adopted by several companies is to use a radial shaft seal with a blocking device such as a spring\cite{shaftspring}. The shaft seal is comprised by a plastic ring that is compressed using a spring. The shaft seal, however, is designed to be leak tight in presence of a fluid such as the lubricant of a shaft. When using it in a sliding seal assembly there is no lubricant and it can easily leak. To avoid leakage we have designed an O-Ring assembly comprised of a O-ring clamped using a stainless steel ring. The ring includes bolts to allow to fix the refrigerator to the cryostat and avoid motion if an overpressure is built up in the Dewar. It also includes an assembly that allows fixing the position of the insert when a overpressure is built up in the Dewar or when the Dewar is pumped.

Other solutions include the use of bellows and a seal assembly\cite{bellows1,bellows2}. These allow for lateral and rocking motions appearing during transport of a large cryostat assembly. In any event, these occupy considerably more space, hampering the usual operation when the cryostat is cold.

The bottom part of a dilution refrigerator contains the experiment and is located inside a vacuum can, often termed the inner vacuum chamber, as opposed to the 'outer' vacuum chamber of the containing Dewar. The inner vacuum chamber is most often made in stainless steel and is partly immersed into liquid Helium. The vacuum chamber connects the upper flange of the chamber to the liquid Helium bath. The upper flange needs usually to be as cold as possible, to facilitate condensation of cryogens that arrive from room temperature through tubing. To this end, one solution is to weld a copper tube to the outside of the stainless steel inner vacuum chamber. This considerably increases the weight of the inner vacuum chamber and makes its cooling from room temperature costly in terms of evaporated liquid Helium.

To reduce the weight of the inner vacuum chamber, we have built a chamber out of Aluminum. The chamber comprises an indium seal assembly on the top to connect the chamber to the dilution insert. The chamber also comprises an indium seal on its bottom. The weight of the chamber is reduced by a factor of three, which correspondingly reduces the cryogenic liquid required for cooling or decreases the time needed to cool down again. The thermal conductivity of Aluminum alloy is about an order of magnitude higher than the thermal conductivity of stainless steel. This allows for effective cooling of the upper flange.

\subsection{Acknowledgments}

Work was supported through grant numbers FIS2017-84330-R, MDM-2014-0377 and RYC-2014-16626 of AEI, by the Comunidad de Madrid through program Nanofrontmag-CM (S2013/MIT-2850), by the European Research Council PNICTEYES grant agreement no. 679080 and by EU COST CA16218 Nanocohybri. Authors acknowledge the staff of segainvex at UAM.

\section{Design files}

\subsection{Design Files Summary}

\tabulinesep=1ex
\begin{tabu} to \linewidth {|X|X|X[1.5,1]|X[1.5,1]|}
\hline
\textbf{Design filename} & \textbf{File type} & \textbf{Open source license} & \textbf{Location of the file} \\\hline
Design file 1 & Figure & CERN OHL & Available with the article, see Fig.\ref{Fig1}  \\\hline
DesignFile2 CAD and DesignFile2 STL & CAD plots & CERN OHL & https://osf.io/e6k7r/ \\\hline

DesignFile3 CAD and DesignFile3 STL & CAD plots & CERN OHL & https://osf.io/e6k7r/ \\\hline

\end{tabu}\\

Design file 1: Connecting schemes for the manual pump.

Design file 2: Design files, including all details, of the heat exchanger, of the scheme of the modifications of the compressor and of the parts made for the compressor.

Design file 3: Design files, including all details, of the sliding seal assembly and of the inner vacuum chamber.

\section{Bill of materials}

\tabulinesep=1ex
\begin{tabu} to \linewidth {|X|X|X|X|X|X|X|}
\hline
\textbf{Designator} & \textbf{Component} & \textbf{Number} & \textbf{Cost per unit currency} & \textbf{Total cost} & \textbf{Source of materials} & \textbf{Material type} \\\hline

1 & Manual pump & 1 & 20\$ & 20\$ & Usual sport or camping market & manual pump with 5 liter volume at each stroke and entry and exit ports \\\hline
2 & Rubber tubes and fittings & 2 & 5\$ & 10\$ & Tubes for the recovery &-- \\\hline
\end{tabu}

\begin{tabu} to \linewidth {|X|X|X|X|X|X|X|}
\hline
\textbf{Designator} & \textbf{Component} & \textbf{Number} & \textbf{Cost per unit currency} & \textbf{Total cost} & \textbf{Source of materials} & \textbf{Material type} \\\hline

1 & Compressor Stayer 70 & 1 & 280\$ & 280\$ & Building market & High pressure machine \\\hline
2 & Solenoid valve SMC position normally open & 2 & 50\$ & 100\$ & Compressed air products & Valves \\\hline
3 & Presostat & 1 & 50\$ & 50\$ & Compressed air products & Sensor and valves \\\hline
4 & Balloon & 1 & 1000\$ & 1000\$ & Balloon made of quality rubber & Gas recipient \\\hline
5 & Fittings and pipes & -- & 70\$ & 70\$ & -- & -- \\\hline
\end{tabu}

\begin{tabu} to \linewidth {|X|X|X|X|X|X|X|}
\hline
\textbf{Designator} & \textbf{Component} & \textbf{Number} & \textbf{Cost per unit currency} & \textbf{Total cost} & \textbf{Source of materials} & \textbf{Material type} \\\hline

1 & Machining hardware & 1 & 200\$ & 200\$ & Hardware to machine the shown items & -- \\\hline
\end{tabu}

\section{Build instructions}

\subsection{Manual pump}

\begin{itemize}
\item Remove grease from the entry and exit ports of the manual pump. 
\item Use Araldite to make the pressure reading leak tight, putting epoxy on the juncture between the manometer and the tube. 
\item Use Araldite or other hard epoxy to glue connectors for flexible rubber tubes. 
\item Connect the entry of the manual pump to the recovery line and the exit to the transport Dewar. 
\item When pumping, take care not to build an overpressure in the transport Dewar. This is unlikely, as all transport Dewars have adequate safety valves, but it should nevertheless be noted.
\end{itemize}

\subsection{Closed cycle circuit of cold Helium}

\begin{itemize}
\item Substitute PVC tubes of the compressor with copper pipes, $\Phi=6$ mm. Use conical copper fittings and make the required adaptations between tube diameters. 
\item Replace the pressostat of the compressor with a type Danfoss Kp35 060 pressostat or similar, a security vale (8 bar) and a filter with flow controller and valve. 
\item Replace the solenoid valves of the compressor with valves that are not leaking. Use SMC normally open valves. Be careful not to substitute the one-way valve at the compressor. 
\item Block the draining system of the tank with a screw. 
\item Seal with Araldite or similar all threads of the compressor.
\item Connect the two entry ports of the compressor to a single line using a T.
\end{itemize}

\subsection{Sliding seal assembly}

\begin{itemize}
\item Make the hardware as shown in the design files. 
\item Screw the IVC using Indium seal as usual. Be careful and maintain a smooth aluminum surface to avoid leaks. 
\end{itemize}

\section{Operation instructions}

\begin{itemize}
\item Usual precautions when using high pressure equipment. The manual pump does not require specific precautions. The compressor requires proper setting of security valves.
\item When inserting the cryostat into liquid Helium always hold the cryostat by a thread so that it is not pushed out in presence of overpressure at the Helium recovery lines. 
\end{itemize}

\section{Validation and characterization}

\begin{figure}[ht]
\includegraphics[angle=0,width=\textwidth,clip]{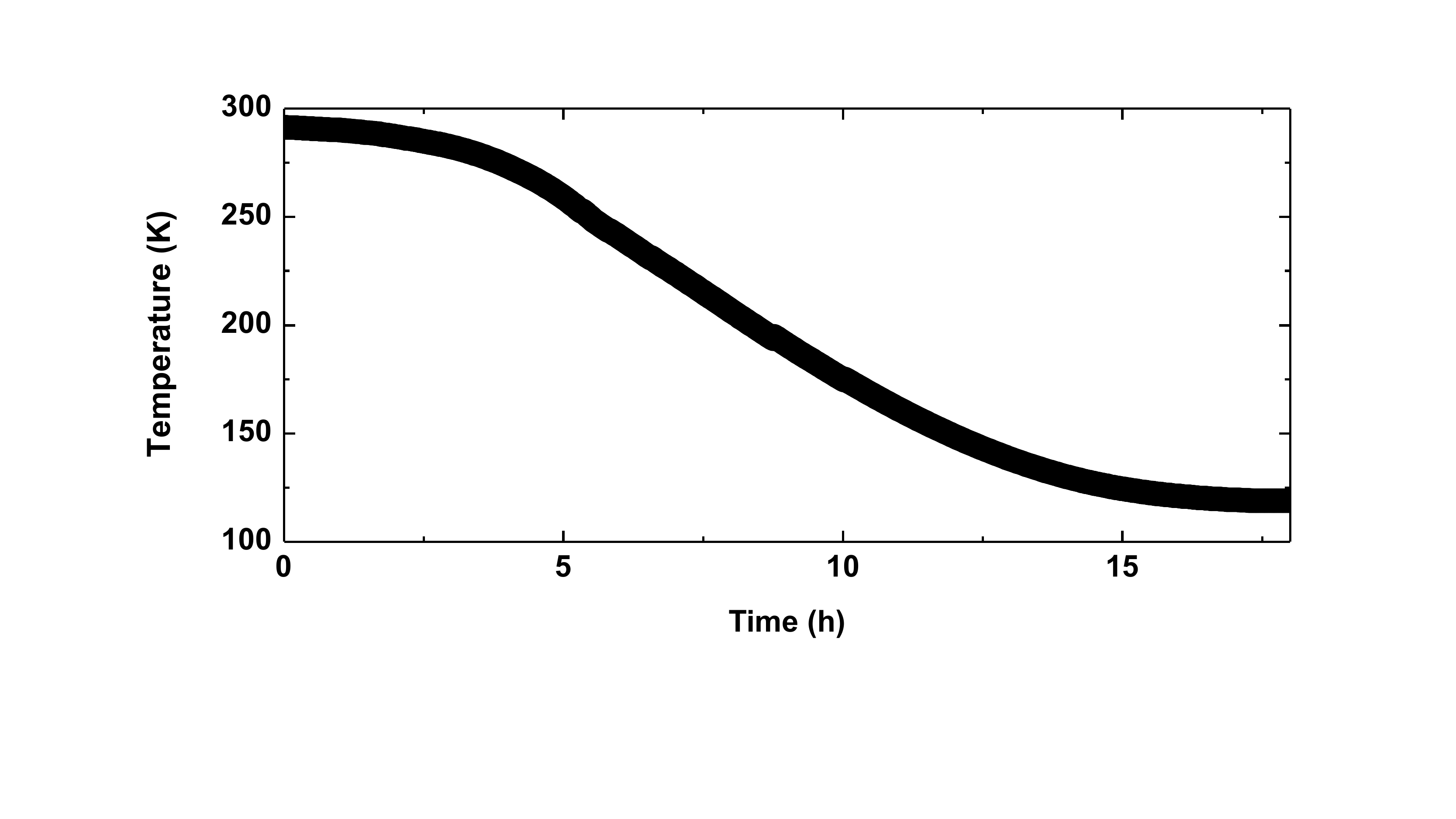}
\vskip -0cm
\caption{Temperature versus time taken during an overnight cool down of a 10 T superconducting magnet using the cold Helium closed-cycle system. During the morning, Helium was transferred with the same evaporation rate as found when using liquid nitrogen in the cryostat.} \label{Fig5}
\end{figure}

\begin{itemize}
\item We have observed that we can have transfer rates as high as 100 liters an hour operating a few times the manual pump. When the Helium transfer Dewar is nearly empty, the pump must be operated a few more times, as the volume needed to fill is much larger. We have used the manual pump in 50 liter and 100 liter transport Dewars. We do not expect major changes when using it in a 200 liter transport Dewar. 
\item In Fig.\ref{Fig5} we show a typical cool down procedure to liquid nitrogen using the closed-cycle system. A bottle of about 60 liters of liquid nitrogen was emptied during the night. 
\item We have observed a decrease of the time needed to cool a cryostat by a factor of three when using the Aluminum inner vacuum chamber. In addition, the time needed to condense the mixture is reduced by a factor of two with respect to the use of stainless steel inner vacuum chambers. 
\end{itemize}

\section{Declaration of interest}
Declarations of interest: none.

\section{Human and animal rights}
Does not apply.


\bibliographystyle{unsrt}

\end{document}